\def\astroph{1}

\ifnum\astroph=0
\documentclass[12pt,preprint]{aastex}       
\usepackage{epsfig,amsmath,natbib}

\graphicspath{{ApJ1/}}
 
\else
\documentclass{emulateapj}
\usepackage{apjfonts}
\usepackage{epsfig,amsmath,natbib,graphicx}

\fi

\newcommand {\ha} {H$\alpha$\,\,}
\newcommand {\hI} {\mbox{H{\sc i}}\,\,}
\newcommand {\hII} {\mbox{H{\textsc{ii}}}\,\,}
\newcommand {\kms} {\,km\,s$^{-1}$\,}
\newcommand {\M} {\mbox{${\cal M}$}}
\newcommand {\mhI} {\M$_{HI}$\,}
\newcommand {\msol} {\M$_\odot$\,}

\newcommand{\fantomm}{{\texttt {\textsc{FaNTOmM}}}}

\slugcomment{Accepted in ApJ Letters}

\shorttitle{\hI rotation curve of M31}
 
\shortauthors{Carignan, Chemin, Huchtmeier \& Lockman}

\begin{document}

\title{Extended \hI Rotation Curve and Mass Distribution of M31}
\author{Claude Carignan\altaffilmark{1}, Laurent Chemin\altaffilmark{1,2},
Walter K. Huchtmeier\altaffilmark{3,4}, and Felix J. Lockman\altaffilmark{5}}

\altaffiltext{1}{Laboratoire d'Astrophysique Exp\'erimentale (LAE),
    Observatoire du mont M\'egantic, and D\'epartement de physique,
    Universit\'e de Montr\'eal, C.P. 6128, Succ. Centre-Ville, 
    Montr\'eal, Qc, Canada H3C 3J7}
 \email{Claude.Carignan@Umontreal.ca}
\altaffiltext{2}{Observatoire de Paris, section Meudon, GEPI, CRNS-UMR 8111 \& Universit\'e Paris 7, 
5 Pl. Janssen, 92195, Meudon, France}
\altaffiltext{3}{Max-Planck-Institut f\"ur Radioastronomie,
    Auf dem H\"ugel 69, Bonn 53121, Germany}
\altaffiltext{4}{Some of the observations reported here were obtained
    with the 100 m telescope of the Max-Planck-Institut f\"ur
    Radioastronomie at Effelsberg}
\altaffiltext{5}{National Radio Astronomy Observatory, P.O. Box 2, 
    Green Bank, WV 24944, USA.  The National Radio Astronomy 
    Observatory is operated by Associated Universities, Inc., under a 
    cooperative agreement with the National Science Foundation.}

\begin{abstract}
New \hI observations of Messier 31 (M31) obtained with the Effelsberg
and Green Bank 100-m telescopes make it possible to measure the
rotation curve of that galaxy out to $\sim$ 35 kpc. Between 20 and 35
kpc, the rotation curve is nearly flat at a velocity of $\sim$ 226
\kms. A model of the mass distribution shows that at the last observed
velocity point, the minimum dark--to--luminous mass ratio is $\sim
0.5$ for a total mass of $3.4 \times 10^{11}$ \msol at $R <$ 35
kpc. This can be compared to the estimated MW mass of $4.9 \times
10^{11}$ \msol for $R <$ 50 kpc.
\end{abstract}


\section{Introduction}

Over the last 35 years, accurate rotation curves (RC) 
have been derived for a large number of galaxies in the Local 
Universe, using  \hI or CO at radio wavelengths,  
or emission lines such as \ha in 
the optical (see Sofue \& Rubin 2001, and references therein). 
With larger telescopes and new instrumentation, RCs have even
been measured for galaxies at redshifts z $\ga 0.5$ 
\citep{vo97,zi03,flo04,bam05}.

Surprisingly, the two nearest massive galaxies, the Milky Way (MW) and 
M31 (at an adopted distance of 780 kpc from McConnachie et al. 2005),
have very poorly defined RCs. 
Our position inside the Milky Way's disk makes it
very difficult to interpret the \hI outside the solar radius, while for M31, the
gas on the receding side is at the same velocity as MW gas 
and it is difficult to disentangle the two components. 
As importantly, the very large angular size of this galaxy
makes it a difficult target for synthesis telescopes (like the VLA)
which are not sensitive to emission from large angular scales.

In spite of numerous attempts to determine the dynamical mass of M31,
a galaxy rich in globular clusters, satellites, \hII regions,
planetary nebulae and having an extended \hI disk 
\citep{har74,gun75,roo79,vdb81,hod92,cou99,mer03,iba04}, 
mass estimates are quite far from satisfactory.  Indeed, we are still
uncertain as to whether M31 or the MW is the most massive member of the
Local Group \citep{eea00}. The kinematics of M31 was studied
extensively about 20 - 30 years ago in \hI
\citep{rob75,eme76,cra80,unw83,brs84} and \ha 
\citep{rub70,bou87,ken89}.
More recent \hI synthesis work was done by \citet{bra90} and collaborators
using the VLA, WSRT and the GBT \citep{thi04}.

Modern 21 cm \hI receivers are up to an order-of-magnitude more
sensitive than those used in the earlier surveys, and with 100-m
telescopes such as those at Effelsberg and Green Bank, it is now
possible to trace the \hI in M31 to a significantly larger radius than
before.  We have thus begun a program of observing 21 cm emission from
the outer parts of this galaxy at $R \geq 90'$ along its major axis.
To avoid confusion with Galactic \mbox{H{\sc i}}, only the approaching
(SW) side of the galaxy was observed.  The major-axis of M31 has a
position angle of 38\degr\ and its central position given in the
Nasa/Ipac Extragalactic Database is $\alpha_{\rm J2000}=
00^h42^m44\fs3$, $\delta_{\rm J2000}=+41\degr16\arcmin09\arcsec$. 
Throughout the article, line-of-sight velocities are given in the 
heliocentric rest frame.

\section{Observations}
\label{sec:obs}  
The first observations were obtained with the Effelsberg 100-m
telescope which has a HPBW = $9\farcm3$.  Spectra were measured every
$4\farcm5$ along the major axis of the SW (approaching) side,
beginning at $R = 90'$, or about 20 kpc from the galaxy's center.
Spectra were measured modulating between the signal band and a
comparison band offset 1.56 MHz, or about -300 km s$^{-1}$.  The
two-channel HEMT receiver was followed by a 1024 channel
autocorrelator split into two bands of 512 channels, with a channel
spacing of 0.64 km/s.  The system temperature was $< 30$ K. Both
polarizations were averaged to reduce the noise.

The brightest \hI emission in M31 comes from a broad, ring-like
structure in the inner parts of the galaxy \citep{eme76,unw83,bra90},
and it is possible that some of this emission might be picked up in a
telescope sidelobe \citep{rei78} and mask any weak \hI signal from the
outer parts of the galaxy.  Care was thus taken during the Effelsberg
observations to ensure that the main sidelobes in the elevation
direction were not falling on the galaxy.  Most of the observations
were done at parallactic angles $\sim$ --45\degr\ to --55\degr, which
is $\sim$ 90\degr\ from the M31 major axis.

\begin{figure*}
\begin{center}
\includegraphics[width=18cm]{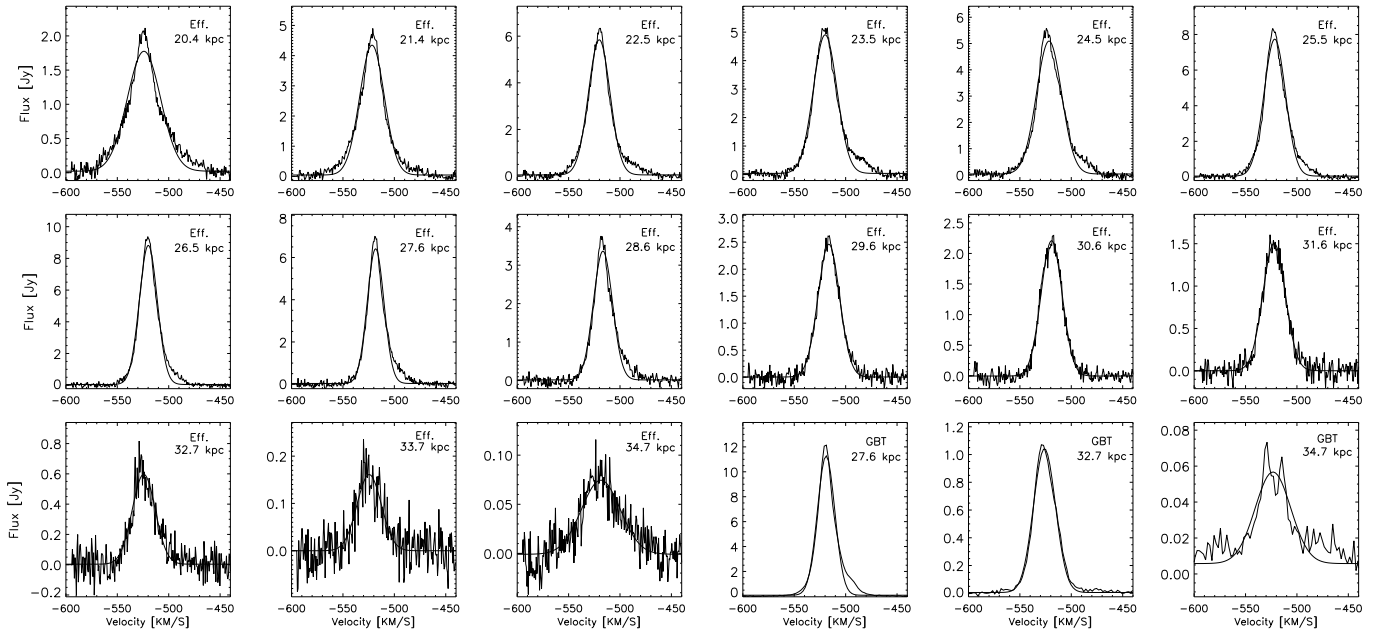}
\caption{\hI spectra -- baseline subtracted -- from the Effelsberg
and GBT 100-m radiotelescopes, with their gaussian fits.  The radius
is given in the top-right corner.  These are only the central portions
of the observed spectra; we believe that all the emission shown here,
including the line wings, is real (cf.~the Effelsberg and GBT profiles
at 27.6 kpc).}
\label{fig1}
\end{center}
\end{figure*}

Sidelobes are less of a problem with the 100-m Green Bank Telescope
(GBT), whose offset optics give it a very clean main beam with a first
sidelobe level more than 25 dB below the peak. The GBT's gregorian
21 cm receiver system overilluminates the subreflector somewhat,
producing a broad forward spillover lobe which contains about 4\% of
the telescope's total response.  However, this sidelobe has a diameter
$> 30\degr$ on the sky and its effects on the current observations
should be negligible \citep{loc05}.

The GBT was used to measure \hI spectra at ten positions along the SW
major axis of M31 in August and September 2005.  At 21 cm, the GBT has
an FWHM angular resolution of $9\farcm1$. Spectra were taken with 2.5
\kms effective velocity resolution over a range $> 1000$ \kms centered
on the M31 velocity. Frequency switching `in band' gave good
instrumental baselines and good sensitivity.  The dual--polarization
receiver had a system temperature T$_{\rm sys} \sim$ 18 K. The ten
positions were observed for times varying from a few minutes to more
than one hour, with a typical value being 40 minutes.

\hI emission was detected out to $R =45$ kpc with the Effelsberg dish,
but here we will discuss only data at  $\leq35$ kpc where spectra were 
measured with both Effelsberg and the GBT. Moreover, the \hI disk 
of M31 is known to be 
slightly warped (e.g. Newton \& Emerson 1977; Henderson 1979). 
Knowing our beamsize and estimates of the warp parameters, we have calculated  that \hI profiles obtained 
along a constant line of 38\degr\ are reliable only up to 35 kpc. 

Within this range the agreement between data from the two telescopes is good.  
For the positions observed with both 
telescopes (27.6, 32.7, 34.7 kpc), 
the mean difference in radial velocity is less than 2.5 \kms, and 
at these three locations the radial velocity is thus given by the 
average of the two observations.
The \hI spectra, baseline subtracted,
with the best fit Gaussian profile superposed, are shown in Figure~\ref{fig1}.
While there is some scaling differences between the profiles coming
from the two telescopes, the profile shapes are remarkably similar (e.g., the
two spectra at $R=27.6$ kpc).

\section{Rotation Curve of M31}
\label{sec:rotcurve}
There are discrepancies between the different determinations of the RC
of M31 in the inner 10 kpc \citep{sof81}, but there is general
agreement for $10 \leq R \leq 30$ kpc. One interesting feature of the
\hI RC of M31 is that it seems to be declining more or less regularly
from the center all the way out to 30 kpc \citep{bra91}. While there
are a few galaxies with declining RCs over a limited radius range
\citep{cap90,cas91,hon97}, this is quite unlike 
what is seen in most spirals, and is one of the
reasons we wished to obtain the velocity information at larger radii 
to check whether that trend continues.

\begin{table}[!t]
\centering
\caption{\hI rotation velocities computed from data
in \citet{unw83} for the inner parts ($R \le 90'$) and from the Effelsberg and 
GBT single dish observations for the outer parts ($R > 90'$).}
\label{table1}
\begin{tabular}{cccc|cccc}
\hline\hline
Radius &   Radius    & $V_{\rm rot}$   & $\Delta V_{\rm rot}$ & Radius &   Radius    & $V_{\rm rot}$   & $\Delta V_{\rm rot}$\\
(1) &  (2) & (3) & (3) & (1) & (2) & (3) & (3)\\
\hline
25.0  &   5.68  &   235.5   & 17.8 & 94.5  &   21.45 &	227.6	& 28.8 \\
30.0  &   6.81  &   242.9   & 0.8  & 99.0  &   22.47 &	226.0	& 28.8 \\
35.0  &   7.95  &   251.1   & 0.7  & 103.5 &   23.50 &	225.7	& 28.8 \\
40.0  &   9.08  &   262.0   & 2.1  & 108.0 &   24.52 &	227.5	& 28.8 \\
45.0  &   10.22 &   258.9   & 6.9  & 112.5 &   25.54 &	227.4	& 28.8 \\
50.0  &   11.35 &   255.1   & 5.7  & 117.0 &   26.56 &	225.6	& 28.8 \\
55.0  &   12.49 &   251.8   & 17.1 & 121.5 &   27.58 &	224.4	& 28.8 \\
60.0  &   13.62 &   252.1   & 7.4  & 126.0 &   28.60 &	222.3	& 28.8 \\
65.0  &   14.76 &   251.0   & 18.6 & 130.5 &   29.62 &	222.1	& 28.8 \\
70.0  &   15.89 &   245.5   & 28.8 & 135.0 &   30.65 &	224.9	& 28.8 \\
75.0  &   17.03 &   232.8   & 1.0  & 139.5 &   31.67 &	228.1	& 28.8 \\
80.0  &   18.16 &   232.0   & 14.2 & 144.0 &   32.69 &	231.1	& 28.8 \\
85.0  &   19.30 &   235.7   & 4.6  & 148.5 &   33.71 &	230.4	& 28.8 \\
90.0  &   20.43 &   229.3   & 13.8 & 153.0 &   34.73 &	226.8	& 28.8 \\
\hline
\end{tabular}
 
 Notes on columns: \ (1) and (2) Radius in arcmin and kpc (resp.) for $D$ = 780 kpc \citep{McC05} (1\arcmin = 227 pc) 
\ (3): Velocity in \kms, for a systemic velocity of $-300$ \kms \citep{deV91}.
\end{table} 

Table~\ref{table1} and Figure~\ref{fig2} give our derived \hI RC.  
 For $R \le 90'$ (or $\sim$ 20.5 kpc),
we have redetermined the \hI velocity field using the early 
\hI data-cube of \citet{unw83} 
and recomputed the rotation velocities by fitting a tilted-ring model
to the velocity map. The \textit{rotcur} task (Begeman 1989) of the
GIPSY package (van der Hulst et al. 1992) was used for that purpose.
No velocities are given inside a radius of $25\arcmin$, where the
deficiency of neutral gas introduces large uncertainties, and where it
is probable that there are large non-circular motions.
The error-bars for the inner RC are given by the difference 
between the velocities of the approaching and receding 
sides of the galaxy, except for the very few points where the formal 
error given by \textit{rotcur} is chosen because it 
exceeds this difference.

The inner part of the new RC compares well 
with the composite one presented in Widrow et al. (2003), which was
compiled from several data sets. The curves of the receding 
and approaching halves are shown to illustrate the symmetry 
of the gas motions in the galaxy: except, perhaps, near  $R \sim 70'$ 
 where a difference of $\sim$ 29 \kms\ is observed, similar values are derived for the 
two sides of the galaxy. This asymmetry was also observed in previous \hI data (Braun 1991).  

The points at $R > 90'$ were computed from the new single-dish
spectra, using the same inclination as for the inner parts (77\degr)
and a systemic velocity of $-300$ \kms \citep{deV91}. The associated
errors are taken to be equal to the largest error found in the inner
part of the curve.  The most striking feature of our new curve is
that, far from having a declining RC \citep{bra91}, the velocities
between 20 and 35 kpc are found to remain nearly constant at $\sim$
226 \kms.

Models of the M31 warp at large radii imply a slightly increasing
inclination and decreasing position angle of the major axis as a
function of radius \citep{nee77,brg90}. As a
consequence, use of a constant inclination of 77\degr\ instead of a
higher one at large $R$ could lead to an overestimate of the rotation
velocities, but only at most 2.5\% (or 6 \kms) with respect to the
purely edge-on case.  As for the position angle, if the \hI spectra
are not taken exactly along the major axis, the derived velocities
would be an underestimate of the true rotational velocity.  This
strengthens our conclusion that the RC is not declining.  Finally, the
adopted error of $\pm29$ \kms\ is very conservative, and likely an
overestimate of the  errors  which might arise, e.g., from a bad
choice of the orientation parameters.

Braun's conclusion that the M31 RC was declining at $R > 100'$ was
probably determined by his two points at the greatest $R$, which
were measured very close to the galaxy's minor axis (Tab. 3 and Fig. 8
of Braun 1991). For such an highly inclined galaxy, the outer regions
close to the minor axis suffer from large deprojection
errors. Furthermore, the superposition of several gas orbits along the
line-of-sight due to the external warp is very important close the
minor axis.  These effects may lead to a bad estimate of the
rotational velocity for the last two points of Braun's RC. A constant
velocity at large radii, as the trend we derive, is in agreement with the flat
RC model proposed by \cite{brs84a}.

\section{Mass Distribution of M31}

\begin{figure} 
\centering
\includegraphics[width=\columnwidth]{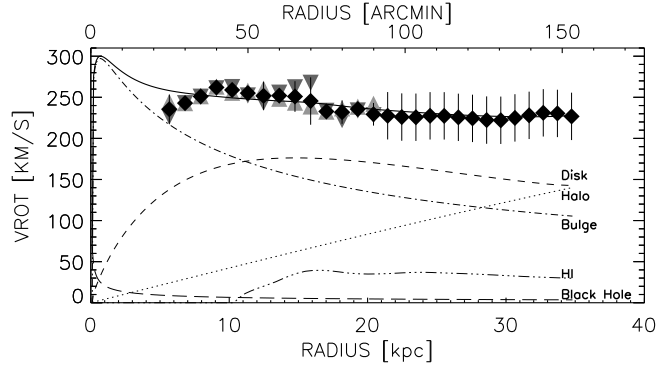}
\caption{Rotation curve and mass model for M31. The new rotation velocities from the Effelsberg
and GBT 100-m  observations are for $R > 21$ kpc. The velocities for $R \leq 21$ kpc are recomputed from the
\citet{unw83} \hI data. The upper pointing triangles (light grey) are for the receding side and 
the bottom pointing (dark grey) ones for the approaching side, as obtained from a tilted-ring 
model (see text). The solid line is the best fit to the data. Each mass component is identified. A
mass of $1.0 \times 10^8$ \msol\ is used for the central  black hole (Bender et al. 2005).}
\label{fig2}
\end{figure}

In order to model the mass distribution of M31 using the RC shown in
Fig.~\ref{fig2}, the luminosity profiles of the luminous components
(bulge \& disk) were derived from the $B$-band parameters given in
Walterbos \& Kennicutt (1987, 1988). The \hI radial profile of
\citet{sof81} was used for the gaseous disk, multiplied by a factor of
4/3 to account for Helium.  The observed colors (0.8 $<$(B-V)$<$ 1.0),
combined with population models, lead to a value for the stellar
component of 2.8 $<$ (M/L$_B$) $<$ 6.5
\citep{bel01}.

Figure~\ref{fig2}  shows the best--fit mass model 
for M31. The model is described in Carignan (1985), and has 
a dark matter (DM) component represented by an isothermal sphere. 
A (M/L$_B$) of 7.5 (in units of M$_\odot$/L$_\odot$) is derived for the 
stellar disk. This is somewhat higher than the value expected
from the colors, which means that this can be considered as a lower limit
for the DM component (maximum disk model). 
Such large (M/L$_B$) values from best--fit models are not uncommon: 
for example,  values of 9.4 \citep{beg89} and 8.5 \citep{bla99} have been found for the galaxy NGC 3198. 
Comparable estimates of (M/L$_B$)  
have been found from measurements of 
the vertical stellar velocity dispersion in the nearly 
face-on galaxies NGC 628 and NGC 1566 (van der Kruit \& Freeman 1984).

For $R <$ 35 kpc, the model gives a stellar mass \M$_{stellar} = 2.3
\times 10^{11}$ \msol, an \hI mass \mhI $= 5.0 \times 10^9$ \msol, and
a dark matter mass of \M$_{DM}= 1.1 \times 10^{11}$ \msol, leading to a
total mass of \M$_{tot} = 3.4 \times 10^{11}$
\msol. This translates in a \M$_{dark}$/\M$_{lum}$ $\sim 0.5$ and a total
dynamical mass--to--light ratio (M/L)$_{dyn} \sim 6$ at the last observed
velocity point.

\section{Conclusions and Future Work}

The main conclusions of this Letter are:
\begin{itemize}
\item The present study has extended the measured rotation curve of M31
out to $\sim 35$ kpc.  Contrary to previous studies (Braun 1991), the
RC does not decline steadily from the center out to the last
velocity point but remains nearly constant at $\sim$ 226 \kms\ for $20
\la R \la 35$ kpc.
\item A total mass for M31 of $3.4 \times 10^{11}$ \msol is derived 
for $R \la 35$ kpc. This is very similar to the mass of $2.8 \times
10^{11}$ \msol (for $R < 31$ kpc) found using kinematical data of
planetary nebulae \citep{eva00}, and can be compared with the MW mass
of $4.9 \pm 1.1 \times 10^{11}$ \msol within 50 kpc \citep{koc96}.
\item A study of the mass distribution gives a lower limit to the 
ratio  \M$_{dark}$/\M$_{lum} \sim 0.5$ at the last observed velocity point. 
This is quite different from the no Dark Matter model of \citet{bra91}.
\end{itemize}

This study is the first part of a program to derive the best possible
rotation curve for M31. Our observations confirm that
\hI can still be detected past 30 kpc and that the outer RC is flat
and not declining. An ideal RC will combine 
the following two data sets: 

(1) \emph{High resolution optical 3--D \ha kinematical data}. 
  These data will be obtained with the
Fabry-Perot system \fantomm\ \citep{gac02,her03}.  With the $\sim 14
\times 14$ arcmin$^2$ field-of-view at the Mont M\'egantic Observatory
telescope, it will be possible to get a mosaic of the whole optical
disk with $\sim$ 20 fields. This will
allow us to get sufficient spatial resolution in the crucial 
rising part of the RC \citep{bla99} and at the same time provide the 2--D
coverage necessary to derive properly the kinematical parameters.

(2) \emph{Arc-minute resolution, high sensitivity synthesis} \hI
\emph{data}. Because of the lack of short spacings, the VLA
observations \citep{bra90} are not sensitive to the large scale \hI
structures expected in M31. However, the DRAO synthesis 
array is sensitive to such structures and retrieves most of the
\hI present in M31. This has proved to be the case for very extended
nearby systems such as NGC 6946 \citep{car90} and DDO 154
\citep{car89}.

These observations are in progress and should provide the optimal kinematical information for 
the best possible study of the mass distribution in M31.

\acknowledgments

We would like to thank O. Hernandez, O. Daigle,
M.-H. Nicol, M.-M. de Denus Baillargeon and D. Naudet for useful discussions and an anonymous referee
for valuable suggestions. LC acknowledges partial support
from the Fonds Qu\'eb\'ecois de Recherche sur la Nature et les Technologies
and CC from the Conseil de Recherches en Sciences Naturelles et en
G\'enie du Canada.

\end{document}